\documentclass [pra,a4paper,twocolumn,showpacs]{revtex4}
\usepackage{dcolumn,bm,amsmath}
\def\veps{\varepsilon}
\def\etal{{\it et al.}\ }
\newcommand{\eref}[1]{(\ref{#1})}
\newcommand{\Eref}[1]{Eq.\,(\ref{#1})}
\def\rtw{\rightarrow}
\begin{document}
\title{Manifestation of the Nuclear Anapole Moment in
            {M1} Transitions in Thallium}
\author{ M.\,G.\,Kozlov}
\affiliation{Petersburg Nuclear Physics Institute, Gatchina,
188300, Russia}
\email{mgk@MF1309.spb.edu}
\date{27 April 2002}

\begin{abstract}
{We calculate nuclear spin-dependent parity non-conserving
$E1$-amplitudes for optical transition $6p_{1/2,F} \rtw 6p_{3/2,F'}$ and
for hyperfine transition $6p_{1/2,F} \rtw 6p_{1/2,F'}$ in $^{205}$Tl.
Experimental limit on the former amplitude placed by Vetter \etal
[PRL,{\bf 74},2658 (1995)] corresponds to the
anapole moment constant $\kappa_a = -0.26 \pm 0.27$.}
\end{abstract}

\pacs{32.80Ys, 11.30.Er, 31.30.Jv}
\maketitle

In 1980 Flambaum and Khriplovich \cite{FK80} pointed out that the nuclear
spin-dependent (NSD) part of the parity non-conserving (PNC) interaction
in heavy atoms is dominated by the contribution of the anapole moment (AM) of
the nucleus \cite{Zel57t}. After that AM was observed in the PNC experiment
with $^{133}$Cs \cite{WBC97}, where the measured value of the AM constant
$\kappa_a$ appeared to be even larger than theoretical prediction for the
``best values'' of the constants of the nuclear PNC interaction (see
\cite{DK02} and references therein). On the other hand, in the most accurate
measurement of the PNC amplitudes $6p_{1/2,F} \rtw 6p_{3/2,F'}$ in $^{205}$Tl
\cite{VMM95}, the NSD amplitude was found to be consistent with zero and
smaller than theoretical predictions \cite{DK02,Khr95}.

In Ref.\,\cite{Khr95} the ratio between NSD amplitude and the dominant nuclear
spin-independent (NSI) PNC amplitude was calculated in the one-particle
approximation. Here we recalculate this ratio using CI+MBPT method
\cite{DFK96a,DFK96b,DKPF98}, which allows to account for both core-valence and
valence-valence correlations. We found that correlation corrections are
relatively large, but do not explain the discrepancy between the measurement
\cite{VMM95} and the theory \cite{DK02}. A more accurate measurement of NSD
amplitude in the optical transition $6p_{1/2,F} \rtw 6p_{3/2,F'}$ is hampered
by the much larger NSI amplitude and by the smallness of the hyperfine
structure of the upper state. Consequently, it may be easier to measure PNC
amplitude for the hyperfine transition $6p_{1/2,F} \rtw 6p_{1/2,F'}$, where
NSI amplitude turns to zero, while NSD amplitude is not suppressed
\cite{NK75t}. Here we find correlation corrections to this amplitude to be
20\%.

In the PNC experiments on the $6p_{1/2} \rtw
6p_{3/2}$ transition in Tl, the ratio:
\begin{equation}\label{i1}
{\cal R} \equiv {\rm Im}(E1_{\rm PNC}/M1)
\end{equation}
of the PNC amplitude to the magnetic amplitude was measured with 1\% accuracy
in Ref.~\cite{VMM95} and with 3\% accuracy in Ref.~\cite{EPBN95}. In those
experiments the hyperfine structure of the lower level $6p_{1/2}$ was
resolved. That allowed to determine ${\cal R}(F)$ for two hyperfine levels
$F=0$ and $F=1$ of the ground state. For the level $F=0$ the only transition
to $F'=1$ of the level $6p_{3/2}$ is allowed, while for the level $F=1$
transitions to both upper hyperfine levels $F'=1,2$ are allowed. Accordingly,
${\cal R}(1)$ is some average for two transitions:
\begin{subequations}
\label{i2}
\begin{eqnarray}
{\cal R}(0) &\equiv& {\cal R}(0,1),
\label{i2a}\\
{\cal R}(1) &\equiv& x^2 {\cal R}(1,1) + (1-x^2) {\cal R}(1,2),
\label{i2b}
\end{eqnarray}
\end{subequations}
where coefficient $x^2$ depends on the intensity of the transitions and on
experimental conditions, such as line width and optical depth.

Observation of the $F$-dependence of the PNC amplitude is important as it
can give information about NSD part of the PNC interaction:
\begin{eqnarray}
     H_{\rm PNC} = H_{\rm NSI} +\! H_{\rm NSD}
      = \frac{G_F}{\sqrt{2}}                             
     \Bigl(-\frac{Q_W}{2} \gamma_5 + \frac{\kappa}{I}
     \vec{\alpha} \vec{I} \Bigr) \rho(\vec{r}),
\label{e1}
\end{eqnarray}
where  $G_F = 2.2225 \times 10^{-14}$ a.u. is the Fermi constant of
the weak interaction, $Q_W$ is the nuclear weak charge, $\kappa$ is
the dimensionless coupling constant, $\gamma_5$ and $\vec{\alpha} \equiv
\gamma_0 \vec{\gamma}$ are the Dirac matrices, $\vec{I}$ is the
nuclear spin ($I=\frac{1}{2}$ for both stable isotopes $^{205}$Tl and
$^{203}$Tl), and $\rho(\vec{r})$ is the nuclear density distribution.

There are three main contributions to the
coupling constant $\kappa$ in NSD part of the PNC
interaction \eref{e1}:
\begin{eqnarray}
     \kappa &=& -\frac{2}{3}\kappa_a +\kappa_2 + \kappa_{Q_W},
\label{e1a}
\end{eqnarray}
where AM contribution is given by the constant $\kappa_a$ \cite{FK80}, the
constant $\kappa_2 = \frac{\lambda}{2}(4\sin^2\theta_W-1)\approx -0.06$
corresponds to the NSD weak neutral currents\,\footnote{Note, that radiative
corrections can change $\kappa_2$ quite noticeably.}. The term $\kappa_{Q_W}$
is induced by the interference of the NSI interaction with the hyperfine
interaction. For heavy nuclei this constant is proportional to $A^{2/3}$
\cite{FK85a,BP91b}, and for Tl $\kappa_{Q_W} \approx 0.02$. Substituting these
numbers in \eref{e1a}, we get:
\begin{eqnarray}
     \kappa = -\frac{2}{3}\left(\kappa_a - 0.06 \right)\,.
\label{e1b}
\end{eqnarray}
Theoretical predictions for AM constant depend on nuclear model and vary
within the range $0.1 \le \kappa_a \le 0.4$ (see \cite{DK02} and references
therein). On the other hand, for a given nuclear model, one can use measured
values of $\kappa_a$ to get information on the coupling constants of the
nuclear $P$-odd interaction \cite{FM97,HLR01,HW01}.

In this article we calculate the NSD amplitudes $6p_{1/2,F}\rtw 6p_{3/2,F'}$
and use \Eref{e1b} and experimental results from \cite{VMM95} to place a limit
on the AM constant $\kappa_a$. Following \cite{VMM95,Khr95} we use
parametrization:
\begin{equation}
\label{i5}
  {\cal R}(F,F') = C(Z) [Q_W - 6 \kappa \xi(F,F')],
\end{equation}
that links NSD amplitude to NSI one via the function $\xi(F,F')$.
According to \eref{i2}, one can define a function $\xi(F)$ as follows:
\begin{subequations}
\label{i7}
\begin{eqnarray}
  \xi(0) &=& \xi(0,1),
\label{i7a}\\
  \xi(1) &=& x^2 \xi(1,1) + (1-x^2) \xi(1,2).
\label{i7b}
\end{eqnarray}
\end{subequations}

An important property of the one-particle approximation is the equality
$\xi(1,1) = \xi(1,2)$ \cite{Khr95}, which means that $\xi(1)$ does not depend
on the coefficient $x^2$ in \eref{i2b} and \eref{i7b}. Numerical values,
obtained in \cite{Khr95}, are:
\begin{equation}
\label{i6}
  \xi_{\rm op}(0) = 0.87; \qquad \xi_{\rm op}(1) = -0.29\,.
\end{equation}
In general, when electron correlations are taken into account, $\xi(1,1) \ne
\xi(1,2)$. Then, one has to use \Eref{i5} for ${\cal R}(F,F')$ and
calculate function $\xi(F',F)$. After that, experimental function $\xi(F)$ is
given by \eref{i7}. Consequently, the separation of NSI and NSD amplitudes
depends on the factor $x^2$.

NSI amplitude was studied many times, the most advanced and accurate
calculations being \cite{DFSS87a,KPJ01} (for earlier references see
\cite{Khr91}). It was shown there that many-body corrections to PNC amplitudes
in Tl can be important. That stimulated us to recalculate function
$\xi(F,F')$. We follow here the same procedure, which was used in
\cite{KPJ01}. It is based on the combination of the many-body perturbation
theory for core-valence correlations and the configuration interaction for
three valence electrons (CI+MBPT method) \cite{DFK96a,DFK96b,DKPF98}.


Most of the technical details of this calculation, such as basis sets,
configuration sets, etc., are the same as in Ref.\,\cite{KPJ01}, where a
number of test calculations were made for the spectrum, hyperfine constants,
$E1$-amplitudes, and polarizabilities. All these parameters were shown to be
in good agreement with the experiment. That allowed to estimate the accuracy
of the calculation of NSI amplitude to be better than 3\%. Here we use the
same wave functions for the states $6p_j$, but neglect several smaller
corrections, such as structural radiation, to the effective operators for
valence electrons. The normalization correction is the same for NDI and NSD
amplitudes and does not affect the function $\xi(F,F')$.

In order to find PNC amplitude we solve inhomogeneous equations:
\begin{eqnarray}
        (E_{6p_{3/2}} - H^{\rm eff}) \Psi_{a,m}^{(D)}
        &=& D^{\rm eff}_z \Psi_{6p_{1/2},m},
\label{pnc2}\\
        (E_{6p_{1/2}} - H^{\rm eff}) \Psi_{b,m}^{(D)}
        &=& D^{\rm eff}_z \Psi_{6p_{3/2},m},
\label{pnc3}
\end{eqnarray}
where $H^{\rm eff}$ is the effective Hamiltonian for valence electrons, which
accounts for core-valence correlations within the second order many-body
perturbation theory \cite{DFK96a,DFK96b}, $D^{\rm eff}_z$ is $z$-component of
the effective $E1$-amplitude in the length-gauge \cite{PRK99b}, and $m$ is
magnetic quantum number. Solutions of these equations can be decomposed in
terms with definite angular quantum number $J$:
\begin{eqnarray}
        \Psi_{i,m}^{(D)}
        &=& \sum_J \Psi_{i,J,m}^{(D)}; \quad i = a,b.
\label{pnc3a}
\end{eqnarray}

NSI amplitude can be found by calculating the following matrix
elements:
\begin{eqnarray}
        &&{E1}_{\rm NSI}
        = (-1)^{\frac{3}{2}-m}
        \left(\begin{array}{ccc}
        \frac{3}{2} & 1 & \frac{1}{2} \\
        -m & 0 & m
        \end{array}\right)^{-1}
\label{pnc5}
        \\ &&\times
        \left(
        \langle \Psi_{6p_{3/2}}|
        H^{\rm eff}_{\rm NSI}|\Psi_{a,3/2}^{(D)}\rangle
        +
        \langle \Psi_{b,1/2}^{(D)}|
        H^{\rm eff}_{\rm NSI}|\Psi_{6p_{1/2}}\rangle
        \right),
\nonumber
\end{eqnarray}
where we skip index $m$ in matrix elements and take advantage of the fact
that $H^{\rm eff}_{\rm NSI}$ is diagonal in quantum number $J$.
NSD part of the PNC interaction \eref{e1} can change this quantum number,
and corresponding amplitudes have more complicated form:
\begin{eqnarray}
        {E1}_{\rm NSD}
        &=& \sum_{J=1/2}^{5/2} C(J,F,F')
        \left(
        \langle \Psi_{6p_{3/2}}|
        H^{\rm eff}_{\rm NSD}|\Psi_{a,J}^{(D)}\rangle\right.
\nonumber
        \\
        &+& \left.
        \langle \Psi_{b,J}^{(D)}|
        H^{\rm eff}_{\rm NSD}|\Psi_{6p_{1/2}}\rangle \right),
\label{pnc6}
\end{eqnarray}
where constants $C(J,F,F')$ are some combinations of the $6j$-coefficients
(see \cite{PK01} for details).

All wave functions in Eqs.\,\eref{pnc5}, \eref{pnc6} are many-electron
ones. In the one-particle approximation, these expressions are
simplified, and both NSI and NSD parts of the PNC amplitude have the form:
\begin{eqnarray}
        {E1}_{\rm PNC}
        =\! \sum_n \!\frac{\langle 6p_{3/2}||D||ns_{1/2}\rangle
        \langle ns_{1/2}|H_{\rm PNC}| 6p_{1/2}\rangle}
        {\veps_{6p_{1/2}}-\veps_{ns_{1/2}}}.
\label{pnc7}
\end{eqnarray}
The sum here runs over occupied ($n=1,\dots,6$) and vacant ($n>6$) states.
Contribution of the occupied states with $n \le 5$ is very small, while $n=6$
contributes almost as much as the whole sum over vacant states. The term $n=6$
corresponds to amplitudes with the index $b$ in \eref{pnc5} and \eref{pnc6}.
It is seen, that all intermediate states in \eref{pnc7} have $J=1/2$. This
leads to the equality $\xi(1,1) = \xi(1,2)$, which is not correct for a more
general case of \Eref{pnc6}. The many-body corrections are strongest for the
weak amplitude $F=1 \rtw F'=1$, which affects the value of $\xi(1,1)$.

\begin{table}[tb]
\caption{Table 1. Calculated values of $\xi(F,F')$ in different approximations:
configuration interaction (CI) for three valence electrons, and CI+MBPT
method; $a$ and $b$ correspond to two contributions in Eqs.\,\eref{pnc5} and
\eref{pnc6}.}

\begin{tabular}{lcccccc}
\hline
\hline
&\multicolumn{3}{c}{CI}
&\multicolumn{3}{c}{CI+MBPT}\\
\multicolumn{1}{c}{$F,F'$}
&\multicolumn{1}{c}{$a$}
&\multicolumn{1}{c}{$b$}
&\multicolumn{1}{c}{total}
&\multicolumn{1}{c}{$a$}
&\multicolumn{1}{c}{$b$}
&\multicolumn{1}{c}{total}\\
\hline
 0, 1 &$  1.09 $&$  1.29 $&$  1.20 $&$  1.08  $&$  1.12 $&$  1.10 $\\
 1, 1 &$ -0.498$&$ -0.513$&$ -0.506$&$ -0.500 $&$ -0.431$&$ -0.462$\\
 1, 2 &$ -0.337$&$ -0.413$&$ -0.378$&$ -0.331 $&$ -0.361$&$ -0.348$\\
\hline
\hline
\end{tabular}
\end{table}

Our results for the function $\xi(F,F')$ are given in Table~1. We find
them from the calculated amplitudes \eref{pnc5} and \eref{pnc6} in two
approximations. At first, we use configuration interaction method for three
valence electrons with conventional operators. Then, we use second order
many-body perturbation theory to construct effective Hamiltonian
$H^{\rm eff}$ and random phase approximation for the effective operators
$D^{\rm eff}_z$ and $H^{\rm eff}_{\rm PNC}$.

It follows from the comparison of Table~1 with the one-particle approximation
\eref{i6}, that correlation effects enhance NSD amplitudes. For the weakest
amplitude $F=1 \rtw F'=1$ the correlation correction exceeds 50\%. For two
other amplitudes correlations are less important, but still account for
20\%~--~25\% enhancement. Valence correlations are larger for the amplitudes
$b$. The dominant contribution to these amplitudes corresponds to the
intermediate states from the configuration $6s 6p^2$, where correlations
between two $p$-electrons are very strong. In contrast to that, the main
contributions to amplitudes $a$ correspond to configurations $6s^2 np$, where
correlations are much weaker.

We showed above, that correlation corrections to NSD amplitudes are rather
large. Moreover, our values of $\xi(1,1)$ and $\xi(1,2)$ noticeably
differ from each other. That leads to the dependence of the experimentally
observed amplitude \eref{i2b} on $x^2$. The value of this parameter depends
on the experimental conditions. In the linear regime, $x^2$ and $1-x^2$ are
proportional to the intensities of the corresponding lines. That gives
$x^2=\frac{1}{6}$ \cite{Khr91}. Actual experiment \cite{VMM95} was done in the
nonlinear regime, when in the center of the line the light was completely
absorbed, and PNC signal was detected only on the wings. In these conditions,
one can expect that $\frac{1}{6} \le x^2 \le \frac{1}{2}$. Below, we will
perform analysis for each of the limiting cases.

If we substitute values from Table 1 to \Eref{i7b}, we get:
\begin{eqnarray}
        \xi(0) = 1.10; \quad
        \xi(1) =
        \left\{
        \begin{array}{lcc}
        -0.367, & \, & x^2 = {1}/{6},\\
        -0.405, & \, & x^2 = {1}/{2}.\\
        \end{array}
        \right.
\label{d1}
\end{eqnarray}
NSI amplitude can be found as weighted average:
\begin{eqnarray}
        {\cal R}_{\rm NSI} &=&
        \frac{\xi(0) {\cal R}(1)-\xi(1) {\cal R}(0)}{\xi(0)-\xi(1)}\,.
\label{d2}
\end{eqnarray}
Experimental difference between ${\cal R}(1)$ and ${\cal R}(0)$ is only about
1\%. Because of that, both values of $\xi(1)$ from \eref{d1} lead to
the same value of ${\cal R}_{\rm NSI} = -14.68 \times 10^{-8}$ in
agreement with the result from \cite{VMM95}.

The difference $\Delta{\cal R} \equiv {\cal R}(1)-{\cal R}(0)$ can be written
as:
\begin{eqnarray}
        \Delta{\cal R}
        &=& 6 \kappa
        \frac{\xi(0)-\xi(1)}{Q_W} {\cal R}_{\rm NSI}
\label{d3}\\
        &=&\!\! -4 (\kappa_a + 0.06)
        \frac{\xi(0)-\xi(1)}{Q_W} {\cal R}_{\rm NSI},
\label{d4}
\end{eqnarray}
where we use relation \eref{e1b} between $\kappa$ and $\kappa_a$. Table~1 and
\Eref{d1} give $\xi(0)-\xi(1)=1.49\pm 0.02$, and substituting the standard
model value $Q_W=-116.7$ \cite{Gro00}, we get:
\begin{eqnarray}
        \Delta{\cal R}
        &=& (0.051 \pm 0.001) (\kappa_a + 0.06)
        {\cal R}_{\rm NSI},
\label{d5}
\end{eqnarray}
where the error bar corresponds to two values of $x^2$ in \eref{d1} and does
not account for the theoretical error, caused by the neglect of the higher
orders of the many-body perturbation theory. The latter was estimated in
\cite{KPJ01} for NSI amplitude to be close to 3\%. Here we neglect the
structural radiation corrections and few other corrections, which can
contribute on the percent level, so we estimate the actual accuracy of
\Eref{d5} to be about 5\%. On this level, the uncertainty in experimental
conditions described by the parameter $x^2$ is negligible.

Using experimental values from \cite{VMM95}:
\begin{eqnarray}
        {\cal R}_{\rm NSI}
        &=& (-14.68 \pm 0.06 \pm 0.16) \times 10^{-8},
\label{d6}\\
        \Delta{\cal R}
        &=& (0.15 \pm 0.13 \pm 0.15) \times 10^{-8},
\label{d7}
\end{eqnarray}
we get the following result for the AM constant:
\begin{eqnarray}
        \kappa_a
        &=& -0.26 \pm 0.27\, .
\label{d8}
\end{eqnarray}
In an independent measurement \cite{EPBN95} of PNC effects in Tl a very close
central value for the parameter $\Delta\cal R$ was obtained, though with a
three times larger uncertainty. If we use \eref{i6} instead of \eref{d1}, we
get $\kappa_a = -0.32 \pm 0.35$. It means, that correlations account for 30\%
corrections and lead to a smaller absolute value of the AM constant. Note,
that in Ref.\,\cite{VMM95} the approximate values $\xi(0)=1$ and
$\xi(1)=-\frac{1}{3}$ were used instead of the more accurate one-particle
values \eref{i6}, and the relation $\kappa=\!-\frac{2}{3}\kappa_a$ was used
instead of \Eref{e1b}.

It was first recognized by Novikov and Khriplovich \cite{NK75t}, that NSD
operator also leads to the $E1$-amplitude between hyperfine sublevels of the
same electronic state. The most interesting in this respect is the hyperfine
transition in the ground state. Such amplitudes were calculated in the
one-particle approximation for Cs and Tl \cite{NK75t} and for K \cite{GEKM88}.
The only many-body calculation was done recently for Fr \cite{PK01}. It is
straitforward to rewrite \Eref{pnc6} for this case, and all calculations are
similar to those for optical transition. The result in a.u. is:
\begin{eqnarray}
        \langle 6p_{1/2},1||E1_{\rm NSD}|| 6p_{1/2},0 \rangle
        = 2.11 \times 10^{-11}\, {\rm i}\, \kappa,
\label{pnc8}
\end{eqnarray}
where we use the same level of approximation, as above and add normalization
correction \cite{KPJ01}. In the one-particle approximation the
$M1$-amplitude for this transition is equal to $-\frac{\alpha}{2 \sqrt{3}}$.
Correlations change this value only at sub-percent level, and we can safely
use it to calculate $\cal R$:
\begin{eqnarray}
        {\cal R}_{\rm hf}(6p_{1/2})
        &=& -1.00 \times 10^{-8}\, \kappa
\label{pnc9}\\
        &=& 0.67 \times 10^{-8} (\kappa_a + 0.06).
\nonumber
\end{eqnarray}
Comparison of this value with the one obtained in \cite{NK75t} shows that
correlations increase the answer by approximately 20\%. Result
\eref{pnc9} can be compared also to the $F=4 \rtw F'=5$ transition in the ground
state $7s$ of $^{211}$Fr, where ${\cal R} = 3.0 \times 10^{-9} \kappa$
\cite{PK01}. Though the $M1$-amplitude for the hyperfine transition in Tl is
significantly weaker than in Fr, it should be much easier to do the
experiment with stable Tl, than with radioactive Fr. Note, that for lighter
Cs, $\cal R$ is an order of magnitude smaller.

We see that electron correlations account for substantial corrections to AM
amplitudes, but do not explain the difference between the experiment
\cite{VMM95} and the nuclear theory prediction that $\kappa_a
=0.25 \pm 0.15$ \cite{DK02}. The experimental accuracy for NSD amplitude
for $6p_{1/2} \rtw 6p_{3/2}$ transition is not high, because this amplitude is
much weaker, than NSI one. Therefore, it may be very interesting to measure
the hyperfine amplitude \eref{pnc8}, where PNC effects are completely
determined by the NSD part of the weak interaction. Note also, that the
frequencies of the hyperfine transitions for two natural isotopes $^{203}$Tl
and $^{205}$Tl differ by 1\% and should be easily resolved. That gives the
possibility to measure AM constants for each of the isotopes.

\smallskip

The author is grateful to V.\,V.\,Flambaum, I.\,B.\,Khrip\-lovich, and
S.\,G.\,Porsev for valuable discussions. This work was supported by RFBR,
grant No 02~-~02~-~16387.


\end{document}